\newcommand{\knet}{\mbox{$K_{\mathrm{net}}$}}
\newcommand{\demot}{\Delta E_{\mathrm{motor}}}
\newcommand{\dedna}{\Delta E_{\mathrm{DNA}}}

\documentclass[rmp,aps,manuscript,groupedaddress]{revtex4}

\usepackage{graphicx}
\usepackage{color}

\begin{document}

\title{Dynamics of Molecular Motors and Polymer Translocation with Sequence Heterogeneity}

\author{Yariv Kafri$^*$, David K. Lubensky$^\#$ and David R. Nelson$^*$}
\affiliation{$^*$ Department of Physics, Harvard University,
Cambridge, MA 02138} \affiliation{$^\#$BioMaPS Institute, Rutgers
University, Piscataway NJ 08854 and Bell Laboratories, Lucent
Technologies, Murray Hill, NJ 07974}

\begin{abstract}
The effect of sequence heterogeneity on polynucleotide
translocation across a pore and on simple models of molecular
motors such as helicases, DNA polymerase/exonuclease and RNA
polymerase is studied in detail. Pore translocation of RNA or DNA
is biased due to the different chemical environments on the two
sides of the membrane, while the molecular motor motion is biased
through a coupling to chemical energy. An externally applied force
can oppose these biases. For both systems we solve lattice models
exactly both {\it with and without disorder}. The models
incorporate explicitly the coupling to the different chemical
environments for polymer translocation and the coupling to the
chemical energy (as well as nucleotide pairing energies) for
molecular motors. Using the exact solutions and general arguments
we show that the heterogeneity leads to anomalous dynamics. Most
notably, over a range of forces around the stall force (or stall
tension for DNA polymerase/exonuclease systems) the displacement
grows sublinearly as $t^\mu$ with $\mu<1$. The range over which
this behavior can be observed experimentally is estimated for
several systems and argued to be detectable for appropriate forces
and buffers. Similar sequence heterogeneity effects may arise in
the packing of viral DNA.
\end{abstract}

\maketitle

\section{Introduction}

The dynamics of many single molecule experiments can be described
in terms of a ``particle'' moving along a one-dimensional
substrate. For example, polymer translocation through a narrow
pore can be parameterized by the number of monomers threaded
through the pore. The motion of molecular motors such as kinesins,
dyneins, myosin, helicase, DNA polymerase, exonuclease and RNA
polymerase can be described by the location of the motor on the
one-dimensional substrate (microtubules, actin filaments, DNA and
mRNA) on which they move. Similarly, the packing of a newly
replicated DNA or RNA in viruses may be described by the molecular
weight of the packed genome. These systems have been a subject of
much experimental
\cite{HowardBook,Wang98,Visscher99,Kasian96,Meller01,Meller03A,Meller03B,Hen00,Smith01,Gijs00,Maier00}
and theoretical attention
\cite{Bhat03,Magnasco93,Prost94,Julicher97,Julicher98,Bus01,Fisher99,Klafter03,Klafter04,Kolo00,Lattanzi01A,Lattanzi01B,Lattanzi02,Sung96,Lubensky99,Zandi03,Muthu01,Chuang02,Anita03}.

Under most conditions the motion of the coordinate describing the
system is biased in one direction. The bias in the case of
molecular motors and packing of newly replicated viral genomes is
due to a chemical process such as ATP (or more generally, NTP)
hydrolysis, while for polymer translocation it can be generated by
the different chemical environments on the two sides of the pore.
For translocating single stranded DNA, such a bias could be
provided by adding, for example RecA \cite{RecA} or other single
stranded binding proteins (which do not pass through the pore) to
the solution on one side of the membrane. Single molecule
experiments allow another source of bias to be introduced into the
system, namely an externally applied force $F$. This has been
done, for example, by attaching a bead to a molecular motor
\cite{Visscher99} or to the end of the genome which is packed into
the viruses \cite{Smith01} and pulling on it using optical
tweezers. Similarly, charged polymers have been translocated using
an externally applied electric field \cite{Meller01}. An
interesting variant on these experiments is the single molecule
measurements of Wuite et al. \cite{Gijs00} on DNA polymerase,
which converts NTPs (nucleotide tri-phosphates) into a ligated
chain of nucleotides via complementary base pairing
\cite{Maier00}. Wuite et al. apply a force $F'$ not to the motor
itself, but instead across the ends of the ssDNA/dsDNA complex to
create a {\it tension} across the substrate on which the molecular
machine operates. Beyond a critical tension $F'_c$ of order of
$40$ pN, the motor goes backward and turns into an exonuclease.
The severe stretching of the backbone of the complementary DNA
strand for $F'>F'_c$ presumably makes further conversion of NTPs
unfavorable and causes removal of nucleotides by the motor to be
favored. Forward and reverse motion of this enzyme are believed to
be associated with different active sites \cite{Doublie1998}.

Most theoretical treatments of these systems have assumed
homogeneous (or at least periodic) systems. Independent of the
microscopic details, such problems can be described at long times
by a random walker moving along a tilted potential or,
equivalently, a biased random walker. For molecular motors such as
kinesins, dyneins or myosins the assumption of homogeneity is
indeed, in most experiments, entirely appropriate. However, in
other cases the motion is along a one-dimensional {\it disordered}
substrate. This is the case, for example, for RNA polymerases,
exonuclease and DNA polymerases, helicases, the motion of
ribosomes along mRNA, the translocation of RNA or DNA through a
pore, and the packing of a viral genome. In all these systems the
one dimensional substrate reflects the heterogeneity of DNA or
RNA, and leads to a modification of the coarse-grained effective
potential in which the random walker describing the system moves.
The potential now depends in a complicated way on the location
along the substrate. Two examples of potential energy landscapes
of particular interest to us here are {\it random energy} and {\it
random forcing} energy landscapes. We define a {\it random energy}
landscape to be any effectively one-dimensional potential with a
mean slope and fluctuations in the value of the potential with a
finite variance about this linear tilt. A {\it random forcing}
energy landscape has an overall mean slope but with energy
fluctuations which are themselves described by a random walk. In
this case, the energy fluctuations about a linear tilt grow as the
square root of the distance along the substrate. These two types
of energy landscapes have been studied in detail in the
statistical mechanics literature~\cite{Derrida83,Bouchaud90} and
lead to strikingly different long time dynamics. In particular the
random forcing energy landscape leads to behavior quite different
from diffusion with drift when the overall tilt of the landscape
is small, as discussed in detail below.

Recently, the effect of disorder in the form of defect sites in a
rachet model which locally reverse the bias of molecular motors
has been considered \cite{Harms97}, using the methods of
\cite{Julicher97}. It was suggested that even though fluctuations
in the microscopic potential are bounded, the resulting {\it
effective} energy landscape is random forcing. Specifically, it
was argued that when the defect concentration was large enough
anomalous random force dynamics would appear. As pointed out in
\cite{Lubensky02} heterogeneity in base pairing energies also
leads to a random force landscape in the context of DNA unzipping.

In this paper we study the effect of sequence heterogeneity in
both polymer translocation and molecular motors in detail for an
exactly solvable class of simple lattice models. We consider both
systems in the context of single molecule experiments that apply
an external force pulling back on the polymer or the motor, which
in the absence of this force are biased to move in one direction.
We introduce microscopic models for both systems which can be
solved {\it exactly} both with and without disorder. A
generalization of our motor model, discussed in Section V and
Appendix D, can also be used as a very simple model of the DNA
polymerase/exonuclease experiments of Ref. \cite{Gijs00}. One can
also consider closely related models of the packing of a viral
genome. In this case there is an extra source of bias due to the
energetic cost of packing the DNA inside the virus. The externally
applied force acts in conjunction with this bias while the motor
acts against both. The details are very similar to the cases
discussed here, with the exception that the energy cost of forcing
the DNA into the capsid does not necessarily vary strictly
linearly with the amount of DNA that has entered. We do not
include a separate discussion of this interesting system.

\begin{figure}
\includegraphics[width=8cm]{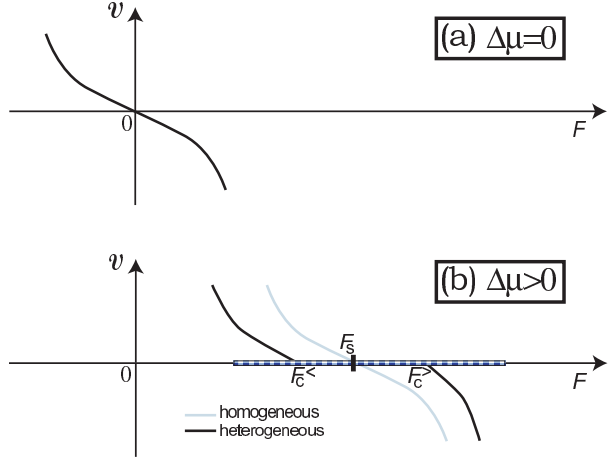} \caption{Schematic
behavior of the drift velocity at long times for homogeneous and
heterogeneous systems as a function of the applied force, where a
positive force {\it resists} the chemically favored direction of
motion. It is assumed that chemical forces (such as ATP hydrolysis
or chemical binding on one side of a pore) lead to a positive
velocity in the absence of a force. (a) No externally applied
chemical bias ($\Delta \mu = 0$). (b) A finite chemical bias
($\Delta \mu > 0$), where the light line corresponds to
homogeneous or periodic environments and the solid line refers to
heterogeneous environments. The anomalous dynamics ($\langle x(t)
\rangle \sim t^\mu$, with $\mu<1$) arises in the vicinity of what
would be the stall force, $F_s$, for the homogeneous system. For
$F_c^< < F < F_c^>$, the effective velocity depends on the width
of the time averaging window, and tends to zero as the width of
the window goes to infinity. The striped line denotes the region
where anomalous diffusion is also present. \label{fig:Phasedia}}
\end{figure}

We show that sequence heterogeneity of single stranded DNA or RNA
and heterogeneous base pairing energies have a dramatic effect on
the dynamics of both systems. For a homogeneous substrate and no
chemical bias, the average velocity changes monotonically through
zero as the external force is varied, changing sign as the force
reverses direction (see Fig. \ref{fig:Phasedia}(a)). When a
chemical bias (which we take to act in the direction opposing the
force) is present, the scenario is similar with the velocity
changing sign at a stall force, $F_s$, which depends on the degree
of chemical bias (see Fig. \ref{fig:Phasedia}(b)). In contrast,
the combination of a disordered substrate and a chemical bias
produces very different behavior for both systems. In this case we
show that generically disorder introduces a random forcing
effective energy landscape which is responsible for the anomalous
dynamics. Similar to the observation of Harms and Lipowsky
\cite{Harms97} a random forcing landscape is generated even if we
neglect an explicit contribution \cite{Lubensky02} from random
base pairing energies. We discuss three different dynamical
regimes which arise due to this landscape as the externally
applied force is varied. The most notable transition arises in the
velocity of the random walker describing the system. Specifically,
we find that there are critical values of the force $F_c^>$ and
$F_c^<$ such that for any force between these values the velocity
is zero in the sense that the average particle position $\langle
x(t) \rangle$, where $\langle \ldots \rangle$ denotes an average
over thermal fluctuations, increases as a {\it sublinear} power of
time. We also discuss an even broader range of forces where the
diffusion is anomalous (see Fig. \ref{fig:Phasedia}(b)). The
transition points between the different types of long-time
dynamics can be calculated exactly for the simple models studied
here.

\begin{figure}
\includegraphics[width=8cm]{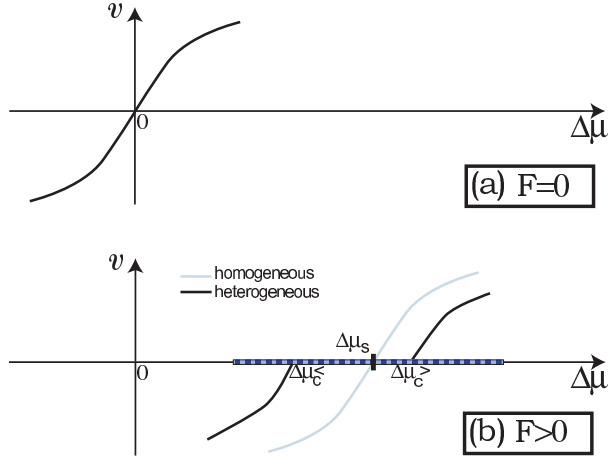} \caption{Schematic
behavior of velocity for homogeneous and heterogeneous systems as
a function of the chemical bias $\Delta \mu$. (a) No externally
applied force. (b) A finite externally applied force for
homogeneous (light line) and heterogeneous (solid line)
substrates. The anomalous dynamics arises in the vicinity of what
would be the stall chemical bias, $\Delta \mu_s$, for the
homogeneous system. As in Fig. \ref{fig:Phasedia}, the striped
line denotes the region where anomalous diffusion is present.
\label{fig:changechem}}
\end{figure}

Under special conditions a random energy landscape is also
possible. In this case the expected behavior as a function of
force is similar to a homogeneous system: The potential
fluctuations simply renormalize the drift velocity and diffusion
constant at long times. That is, as the applied force is varied
the behavior is similar to that of a homogeneous system with no
chemical bias. Provided that random contributions to the energy
landscape not associated with simple conversion of chemical energy
can be neglected, random energy models describe the dynamics in
the absence of chemical bias (see Fig. \ref{fig:Phasedia}(a)) on
heterogeneous substrates.

\begin{figure}
\includegraphics[width=8cm]{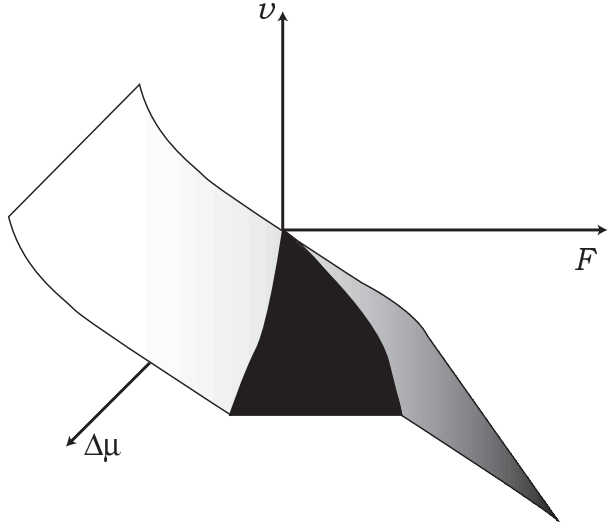} \caption{The dependence of the velocity on
the chemical bias $\Delta \mu$ and the external force $F$. We
neglect for simplicity contributions to a random force landscape
(such as fluctuations in base pairing energies) which may be
present even for $\Delta \mu=0$. Here it is assumed that the
chemical bias always act in the direction opposing the force. The
black wedge denotes a region of sublinear drift with time, i.e.,
effectively zero velocity. \label{fig:BigDia}}
\end{figure}

An alternative way to observe the anomalous dynamics is by holding
the external force constant and varying the chemical bias. This
can be done by changing the concentration of, say, nucleotide
triphosphates for molecular motors, or by changing the
concentration of the polymer binding protein in one chamber for
polymer translocation experiments. In this case, when the force is
held at zero, the velocity changes monotonically in tandem with
the chemical bias (see Fig. \ref{fig:changechem}(a)). However,
when the external force is held constant at a non-zero value, a
region with anomalous dynamics appears as the chemical bias is
varied (see Fig. \ref{fig:changechem}(b)). Between two values of
the chemical bias $\Delta \mu_c^<$ and $\Delta \mu_c^>$, the
displacement of the particle with time is again sublinear, in
contrast to the same experiment performed on a homogeneous
substrate. As illustrated in Fig. \ref{fig:changechem}(b) the
velocity is then a monotonic function of the chemical bias,
changing sign at a stalling chemical bias $\Delta \mu_s$. A
summary of the qualitative behavior of the velocity as function of
both the chemical bias $\Delta \mu$ and the external force $F$ is
shown in Fig. \ref{fig:BigDia}. It is worth noting that there is
no region of sublinear displacement when $\Delta \mu = 0$ because
the energy landscape is then random energy rather than random
forcing, whereas when $F = 0$, there is still a random forcing
landscape everywhere except exactly at stalling, but the
randomness is too small in the vicinity of $\Delta \mu = 0$ to
cause anomalous dynamics.

To keep the discussion simple, Fig. \ref{fig:BigDia} neglects
contributions to a random forcing landscape other than those
produced by the simple conversion of chemical energy along an
inhomogeneous track. Additional random forcing contributions will
arise from, e.g., base pairing energies in the case of helicases,
which open up DNA strands or DNA polymerases and exonucleases,
which add or delete complementary base pairs. Motors, such as RNA
polymerases and ribosomes produce trailing strands of mRNA and
protein respectively. Since these products are themselves
heteropolymers, composed of monomers which interact differently
with the solvent, here too we would expect additional
contributions to a random forcing landscape. Such effects will
only accentuate the anomalous dynamics which is the subject of
this paper.

Before concluding this introduction, we should emphasize our
perspective on the models of polynucleotide translocation and
molecular motors studied here. In an effort to obtain simple,
soluble models which incorporate heterogeneity, we intentionally
neglect important molecular details such as those which describe
the detailed pore interactions of the translocating nucleotides or
distinguish the biological role of motors such as helicases, DNA
polymerase and exonucleases, RNA polymerase, etc. The motors
mentioned above perform important specialized functions such as
opening double stranded DNA, polymerization and depolymerization
or creating messenger RNA while moving along heterogeneous tracks.
Such functions are incorporated into our model simply by adding an
explicit (position-dependent) chemical force to the energy
landscape. More sophisticated attempts to get molecular details
right (see, e.g., Refs. \cite{Anita03},\cite{Simon92} and
\cite{Betterton02}) serve a valuable purpose, which can be
important for modelling some aspects of the dynamics on various
time scales. However, incorporation of sequence heterogeneity,
neglected in most previous modelling efforts, is nevertheless
crucial to correctly describe the anomalous long time dynamics
(e.g., $\langle x (t) \rangle \sim t^\mu$ with $\mu<1$) near the
stall forces in these systems. Otherwise, we expect simple
diffusion with drift (similar to what we find here for homogeneous
models or a random energy landscape) at long times. We do not
expect the multiple intermediate states and numerous rate
constants of more sophisticated models to change our predictions
of heterogeneity-induced anomalous dynamics at long times.

The paper is organized as follows: In Section II, to establish
notation and provide a context for the rest of the paper, we
discuss the homogeneous models for polymer translocation and
molecular motors is some detail. In Section III the effect of the
heterogeneity on the energy landscape is introduced.  Section IV
discusses the resulting dynamical behavior and the exact location
of the transition points within the models. Finally, Section V
estimates the experimental range over which the anomalous dynamics
may be observed for a few representative biological systems and
discusses the effect of finite time experiments on the shape of
the velocity-force curve.

\section{Homogeneous Models}

Before turning to heterogeneous systems we first define
microscopic models for both homogeneous polymer translocation and
molecular motors. The simplicity of both models allows for their
exact solution. Dynamics in heterogenous systems will be treated
in Sections III and IV.

\subsection{Polymer Translocation}

\begin{figure}
\includegraphics[width=7cm]{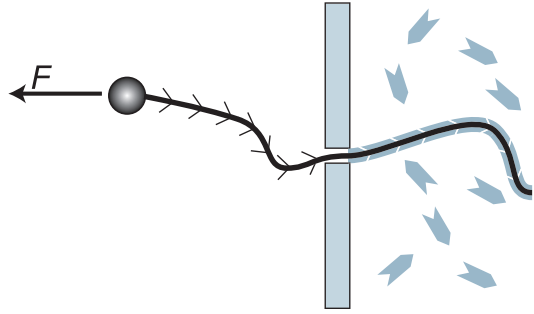} \caption{Schematic
picture of the polymer translocation experimental setup
considered. A polymer is biased to move through the pore by a
solution of binding proteins in the right chamber. A bead exerts a
force in the opposite direction. The arrows reflect the lack  of
inversion symmetry in, e.g., single stranded DNA or RNA.
\label{fig:polymer}}
\end{figure}

An idealized experimental setup is shown schematically in Fig.
\ref{fig:polymer}. A polymer is threading through a narrow pore
located on a two dimensional membrane which separates two
chemically distinct solutions. For concreteness we consider the
right side as containing a polymer binding protein which is absent
in the left hand side. In addition, a bead, through which a
resisting force is exerted on the polymer, is connected to the
left end of the polymer. A model of this kind has been discussed
by P. Nelson \cite{BookPNelson} as a simple example of stochastic
rachet-like dynamics in biological systems (see also
\cite{Peskin93}). Alternatively a force could be applied via an
external electric field acting across the pore on a charged
polymer \cite{Kasian96}.

A convenient representation of the system is through a
one-dimensional random walker located at a coordinate $x$ which
represents the length of the polymer that has translocated to the
righthand side. The conditions under which the full
three-dimensional, multi-species problem can be simplified are
reviewed below. The dynamics of the random walker is governed by
the interaction of the polymer with the pore, the binding of the
protein in the right chamber, and the externally applied force.

\begin{figure}
\includegraphics[width=6cm]{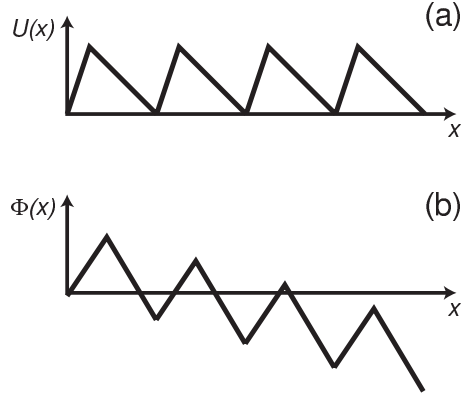} \caption{(a) The periodic
potential due to pore interactions with a translocating polymer
without inversion symmetry. (b) The tilt of this potential
generated by a combination with the binding protein and the
external force. \label{fig:polymerpot}}
\end{figure}

Before turning to a specific microscopic model consider the
general form of the potential experienced by the random walker due
to all these interactions. Because we neglect sequence
heterogeneity in this section, the energy due to interactions with
the pore, $U(x)$, is some periodic function with a period given by
the size of a monomer. An example is the sawtooth or rachet
potential shown in \ref{fig:polymerpot}(a). This type of potential
accounts for an energetic barrier for translocation through the
pore. The lack of inversion symmetry reflects, for example, the
difference in passing single stranded DNA or RNA in the $3' \to
5'$ direction through the pore as opposed to the reverse. The
energy due to the interaction with the polymer binding protein is
however very different and has the form $-F_\mu x$, growing
linearly with $x$. Thus the energy decreases as the polymer
translocates to the righthand side. The value of $F_\mu$ is
governed by the chemical potential difference per monomer, $\Delta
\mu$, of the polymer in the solutions on the righthand and
lefthand sides. This chemical potential difference is a function
of the protein concentration and its binding energy to the polymer
(a more detailed description of $F_\mu$ for the microscopic model
discussed below is presented in Appendix~\ref{chempot}). Finally,
the backward force applied on the bead leads to a contribution to
the energy of the form $Fx$. Upon collecting together these
contributions, the total potential experienced by the random
walker, $\Phi(x)$, is given by
\begin{equation}
\Phi(x)=U(x)-(F_\mu-F)x \;. \label{eq:pot}
\end{equation}
As is evident from the effective energy landscape shown in Fig.
\ref{fig:polymerpot}(b), the random walker is moving in a periodic
potential with an overall slope which depends on the protein
concentration and binding energy as well as the external force.
Such a potential leads on long time scale and large length scales
to motion which is diffusion superimposed on an overall drift
velocity. Thus, the average location of the particle $\langle x
\rangle$ behaves as $\langle x \rangle=vt$ while the mean square
fluctuations about this drift behave as $\langle x^2\rangle -
\langle x \rangle^2 =2Dt$, where $v$ and $D$ depend on $F_\mu-F$
and the details of the rachet potential (see, e.g.,
\cite{Lubensky99}). Here, the brackets $\langle \ldots \rangle$
represent an average over thermal fluctuations.

We emphasize that our simplified description in terms of a single
coordinate $x$ that diffuses and drifts in a one-dimensional
energy landscape is valid only when the translational motion of
the polymer backbone through the pore is the slowest process in
the problem~\cite{Lubensky99}. In particular, this model assumes
that the translocating polymer is not so long that the relaxation
times in the cis (left) or trans (right) chambers exceed the
diffusion time for the backbone through the pore. This simplified
model is also inadequate if the polymer can become bound to the
pore interior for long periods, as recent experiments suggest
occurs for one of the best studied polymer-pore
systems~\cite{Meller03A}. In this case, $x$ will still undergo
biased diffusion on long enough time scales, but its velocity and
diffusion coefficient will no longer be determined by a simple
potential $U(x)$. Finally, the effect of binding proteins can be
captured by a single free energy parameter $\Delta \mu$ only when
their binding and unbinding kinetics are sufficiently fast.  The
opposite limit, in which proteins bind irreversibly, but slowly,
to the polymer, has also received
attention~\cite{Sung96,Simon92,Peskin93}, but we will not consider
it further here.

\begin{figure}
\includegraphics[width=10cm]{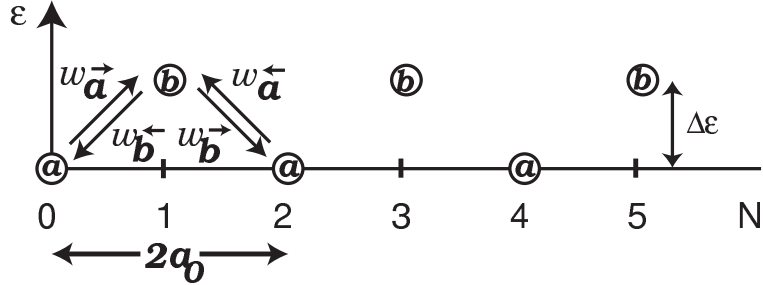} \caption{Graphical representation
of a simplified model for polymer translocation or molecular
motors. These two cases are distinguished by the choice of rate
constants (see text). The distinct even and odd sublattices are
denoted by $a$ and $b$ respectively. \label{fig:motor}}
\end{figure}

We now define a simplified microscopic model for the motion of a
random walker in such a potential. Our model is in the spirit of
those analyzed for motor proteins in \cite{Fisher99,Kolo00} (see
also Sec. II.B), and allows exact results for the diffusion and
drift on long times. In the language of \cite{Fisher99,Kolo00} our
model is a $n=2$ model corresponding to a motor with just two
internal states. More importantly, our model generalizes naturally
to a {\it heterogeneous} version (see Section III) for which exact
results are also possible. We allow $x$ to assume a discrete set
of values $x_m$, where $m=0,1,2 \ldots$ labels distinct $a$ (even)
and $b$ (odd) sites. We can allow different distances between
$x_{m+1}-x_m$ and $x_{m+2}-x_{m+1}$ but require
$x_{m+2}-x_{m}=2a_0$ which we assume for simplicity is the size of
the polymer unit which accommodates a single adsorbed protein. For
a homopolymer the interactions with the pore are some periodic
function with a period which we can take to be $2a_0$. To model
this situation we take odd labelled sites to have a higher energy
than even labelled sites. The arrangement is shown schematically
in Fig. \ref{fig:motor}. Even sites have an energy $\varepsilon=0$
while odd sites (corresponding roughly to the peaks in the rachet
potential of Fig. \ref{fig:polymerpot}) have a higher energy
$\varepsilon=\Delta \varepsilon$. Also, indicated in the figure
are the hopping rates which describe the dynamics of the random
walker. The detailed balance condition (in temperature units such
that $k_B=1$) is satisfied by
\begin{eqnarray}
\label{pore-omegas} w_a^\rightarrow&=& \omega e^{-\Delta
\varepsilon/T-f/2T}
 \nonumber \\
w_b^\leftarrow&=& \omega e^{f/2T}
 \nonumber \\
w_a^\leftarrow &=& \omega' e^{(-\Delta \mu -\Delta
\varepsilon+f/2)/T} \label{eq:polyrates}
\\
w_b^\rightarrow&=& \omega' e^{-f/2T} \;.\nonumber
\end{eqnarray}
Because of the lack of reflection symmetry in the translocating
DNA or RNA (for our model this asymmetry could be represented by
taking $x_1-x_0 \neq x_2-x_1$), we expect the intrinsic hopping
rates to be unequal, $\omega \neq \omega'$. The bias induced by
the interaction of individual monomers with the reservoir of
proteins on one side of the pore has been accounted for by the
chemical potential difference $\Delta \mu$. A more detailed
discussion of the dependence of $\Delta \mu$ on the protein
binding energy and its concentration is given in
Appendix~\ref{chempot}. The effect of the applied force is
included through the parameter $f=Fa_0$. Note that the bias
controlled by $\Delta \mu >0$ arises only for steps from odd to
even sites since a protein is assumed to bind only to a whole
monomer. As pointed out, in \cite{Kolo00}, other $f$-dependences
of the rates consistent with detailed balance are possible. We
shall be content with the simple one displayed in Eq.
\ref{eq:polyrates} which corresponds to choosing $x_1-x_0 = x_2 -
x_1$.

To show that this microscopic model embodies an effective
potential of the form (\ref{eq:pot}), we eliminate the
odd-numbered sites. This elimination can be accomplished by
formally solving the equations of motion for the odd sites,
substituting into the remaining even site equations, and taking he
long-time limit (see Appendix \ref{master}). Alternatively we can
invoke detailed balance and consider an effective energy
difference $\Delta E =E(m+2)-E(m)$ between site $m+2$ and $m$
where $m$ is even. Upon setting
\begin{equation}
\label{eq:db} \frac{W_{m,m+2}}{W_{m+2,m}} \equiv
e^{(E(m+2)-E(m))/T} \;,
\end{equation}
where $W_{n,m}$ is the effective transition rate between site $m$
and $n$, we have
\begin{eqnarray}
\Delta E&=& E(m+2)-E(m) \nonumber \\
&=&T\ln\left( \frac{w_a^\leftarrow w_b^\leftarrow}{w_a^\rightarrow
w_b^\rightarrow}\right) \;. \label{eq:dEdef}
\end{eqnarray}
Use of the rates (\ref{eq:polyrates}) leads to
\begin{equation}
\Delta E=-\Delta \mu+ 2f\label{eq:dEpol}
\end{equation}
as one would expect. Note that when the force vanishes ($f=0$) and
the chemical potential gradient $\Delta \mu =0$ one has $\Delta
E=0$ and no net motion is generated. More generally, an effective
tilted potential of the form (\ref{eq:pot}) is generated, with
$\Delta \mu>0$ causing a drift of the polymer to the right. The
external force on the left can reduce or even reverse the overall
slope. Such a potential inserted into microscopic rate equations
for the even sites (see Appendix \ref{master}) is well known to
lead to diffusion with drift on long time scales and large length
scales.

In fact, for this model using the results of \cite{Derrida83} and
following \cite{Fisher99,Kolo00} one can calculate the velocity
and diffusion constant exactly. After some lengthy calculations,
one obtains for the velocity
\begin{equation}
\label{eq:vel}
v=2 a_0 \frac{w_a^\rightarrow w_b^\rightarrow
-w_a^\leftarrow w_b^\leftarrow}{w_a^\rightarrow
+w_a^\leftarrow+w_b^\rightarrow+w_b^\leftarrow} \;.
\end{equation}
The diffusion constant of the model is given by
\begin{equation}
\label{eq:diff1} D=2 a_0^2 \frac{(w_a^\leftarrow w_b^\leftarrow +
w_a^\rightarrow w_b^\rightarrow)+8w_a^\leftarrow w_b^\leftarrow
w_a^\rightarrow w_b^\rightarrow}{(w_a^\leftarrow + w_b^\leftarrow
+ w_a^\rightarrow + w_b^\rightarrow)^3} K \;,
\end{equation}
with
\begin{eqnarray}
\label{eq:diff2}
 K &=&
(w_a^\leftarrow)^2 + (w_b^\leftarrow)^2 + (w_a^\rightarrow)^2 +
(w_b^\rightarrow)^2 \nonumber \nonumber \\  &+& 2 (w_a^\rightarrow
w_a^\leftarrow + w_b^\rightarrow w_b^\leftarrow+w_a^\leftarrow
w_b^\rightarrow+w_a^\rightarrow w_b^\leftarrow) \;.
\end{eqnarray}

It is interesting to set $f=0$ and consider the limit of $\Delta
\mu /T \ll 1$ (small chemical bias, no external force) and the
limit $\Delta \mu /T \to \infty$ and $\Delta \mu \gg \Delta
\varepsilon$ (large chemical bias, no external force). When
$\Delta \mu /T \ll 1$ the velocity takes the linear response form
\begin{equation}
v=\frac{2 a_0 \omega \omega' e^{-\Delta \varepsilon
/T}}{(\omega+\omega')(1+e^{-\Delta \varepsilon /T})} \frac{\Delta
\mu}{T}
\end{equation}
In the limit of $\Delta \mu /T \to \infty$ and $\Delta \mu \gg
\Delta \varepsilon$, the velocity saturates at $v_{\rm max}$, with
\begin{equation}
v_{\rm max}=2\frac{\omega \omega'}{\omega+ e^{\Delta \varepsilon
/T}(\omega+\omega')} \label{eq:vmaxpol} \;.
\end{equation}
In both cases the velocity is a decreasing function of $\Delta
\varepsilon$, as one might expect because the rate limiting step
in this simple polymer translocation model is the energetic
barrier as each successive segment passes through the pore
potential.

For the diffusion constant one finds similarly in the limit
$\Delta \mu / T \ll 1$
\begin{eqnarray}
D&=&4 a_0^2 \frac{\omega \omega'e^{-\Delta \varepsilon
/T} }{(\omega+\omega')(1+e^{-\Delta \varepsilon /T})} \nonumber \\
&-&2 a_0^2 \frac{ \Delta \mu}{T}\frac{(\omega (1+e^{-\Delta
\varepsilon /T})+\omega' (1-e^{-\Delta \varepsilon
/T}))}{(\omega+\omega')^2(1+e^{-\Delta \varepsilon /T})^2} \;.
\end{eqnarray}
Like the velocity, in this regime the diffusion constant decreases
as $\Delta \varepsilon$ increases. Note that the diffusion
constant deceases when $\Delta \mu$ increases. This behavior
arises since the rate of backward steps decreases as $\Delta \mu$
increases. In the limit $\Delta \mu /T \to \infty$, and $\Delta
\mu \gg \Delta \varepsilon$ we find that the diffusion constant
saturates at
\begin{equation}
D_{\rm max}= a_0^2 \frac{ 2 \omega
\omega'((\omega+\omega')^2e^{2\Delta \varepsilon /T} +\omega^2
(1+2e^{\Delta \varepsilon /T})) }{(\omega+ e^{\Delta \varepsilon
/T}(\omega+\omega'))^3} \;,
\end{equation}
which also decreases with $\Delta \varepsilon$. The diffusion
constant again decreases as a function of $\Delta \varepsilon$ due
to the rate limiting step of the passage through the pore.

\subsection{Molecular Motors}

\begin{figure}
\includegraphics[width=8cm]{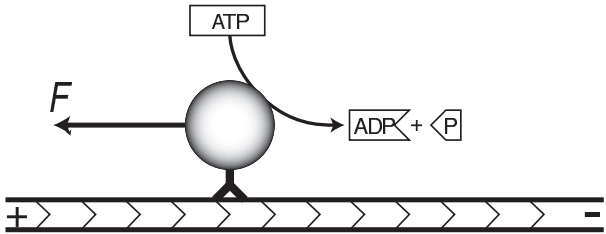} \caption{Setup modelled. The motor is
moving from the $+$ end to the $-$ end. A force is pulling on the
motor in the opposite direction.  Note that some of the specific
biological examples considered in the text are more complicated
and may be driven by energy sources other than ATP.
\label{fig:motorsu}}
\end{figure}

A typical experimental setup is shown in Fig. \ref{fig:motorsu}.
The motor attempts to move from the $+$ end to the $-$ end by
utilizing the chemical energy stored in ATP or some other source
of chemical energy. For RNA polymerase, this energy source would
be the nucleotide triphosphates which are converted into mRNA (not
shown). A force (say from an optical tweezer) pulls in the
opposite direction to the motion generated by the ATP.  In this
section, we focus primarily on models of relatively simple motors
as in Fig.~\ref{fig:motorsu} and mention only in passing more
complicated effects associated with motors such as helicases or
RNAp.

Theoretical models of molecular motors \cite{Julicher97} have
demonstrated how an effective potential of the form (\ref{eq:pot})
is generated as a result of the coupling to an energy source like
ATP for a general class of periodic substrate potentials which
lack inversion symmetry. Here we again introduce a simple model
for a two level rachet which is amendable to an exact solution,
similar to an $n=2$ version of the models of Fisher and Kolomeisky
\cite{Fisher99,Kolo00}. Like the model for polymer translocation
in Section A, this motor model will allow us to study the effect
of heterogeneity. We first consider the homogeneous motor model in
some detail.

We again consider a one-dimensional lattice where even sites have
energy $\varepsilon=0$ while odd sites have an energy
$\varepsilon=\Delta \varepsilon$. The odd sites represent an
``inchworm''-like walking which is facilitated by chemical energy
released by, e.g., hydrolysis of ATP. The transition rates
depicted in Fig. \ref{fig:motor} now take a different form, namely
\begin{eqnarray}
w_a^\rightarrow&=&(\alpha e^{\Delta \mu /T} + \omega)e^{-\Delta
\varepsilon/T-f/2T}
 \nonumber \\
w_b^\leftarrow&=&(\alpha  + \omega)e^{f/2T}
 \nonumber \\
w_a^\leftarrow &=&(\alpha' e^{\Delta \mu /T} + \omega')e^{-\Delta
\varepsilon/T+f/2T} \label{eq:rates}
\\
w_b^\rightarrow&=&(\alpha'  + \omega')e^{-f/2T} \;.  \nonumber
\end{eqnarray}
Note that there are two parallel channels for the transitions
\cite{Julicher97}. The first, represented by contributions
containing $\alpha$ and $\alpha'$, arise from utilization of
chemical energy. The second channel, represented by the terms
containing $\omega$ and $\omega'$, correspond to thermal
transitions unassisted by the chemical energy. $\Delta \mu$ is
given by the standard relation \cite{HowardBook},
\begin{equation}
\Delta \mu =T \left[\ln \left( \frac{[ATP]}{[ADP][P]}\right) -\ln
\left( \frac{[ATP]_{\rm eq}}{[ADP]_{\rm eq}[P]_{\rm eq}} \right)
\right]
\end{equation}
where the square brackets $\left[ \ldots \right]$ denote
concentrations under experimental conditions and the $\left[
\ldots \right]_{\rm eq}$ denote the corresponding concentrations
at equilibrium. We have again assumed the external applied force
$f$ biases the motion in a particularly simple way. If the
substrate lacks inversion symmetry, we have $\alpha' \neq \alpha$
and $\omega' \neq \omega$. As discussed in the introduction, in
some cases an additional force arises from, e.g., base pairing
energies in the case of helicases, DNA polymerase and
exonucleases. Similarly an addition force arises also for motors
such as RNA polymerase and ribosomes which produce trailing
strands of mRNA or protein respectively. Here we ignore such
contributions although they could easily be added in a simple way
to the model through a redefinition of $f$ through $f \to f+f_\mu$
where $f_\mu$ is the additional force. The model is formally
similar to the model of polymer translocation, although the
different functional form of
$w_a^\rightarrow,w_b^\leftarrow,w_a^\leftarrow$ and
$w_b^\rightarrow$ has important consequences.

First we consider the effective energy landscape. To this end, we
again eliminate the odd sites and describe the remaining dynamics
in terms of an effective potential. This is the effective
potential under which a random walker satisfying detailed balance
would exhibit the same dynamics. From a formula similar to Eq.
\ref{eq:db}, ones finds that $\Delta E =E(m+2)-E(m)$, where $m$ is
an even site, is given by
\begin{eqnarray}
\Delta E&=&T\ln\left( \frac{(\alpha + \omega)(\alpha' e^{\Delta
\mu /T} + \omega')}{(\alpha e^{\Delta \mu /T} + \omega)(\alpha' +
\omega')}\right)
\nonumber \\
&+&2f \label{eq:dE}
\end{eqnarray}
where we have used the rates (\ref{eq:rates})

Note that when the external force $f=0$ and the ATP/ADP+P chemical
potential difference $\Delta \mu =0$, one has $\Delta E=0$ and no
net motion is generated. Also, when there is directional symmetry
in the transition rates $\alpha=\alpha'$, $\omega=\omega'$ and
$f=0$ one has $\Delta E=0$ even when $\Delta \mu \neq 0$. Absent
this symmetry, chemical energy can be converted to motion and an
effective tilted potential is generated. Although, these
conditions are equivalent to those presented in
\cite{Prost94,Julicher97} for continuum models, it is interesting
to see them at work in the ``minimal'' model studied here (see
also \cite{Fisher99} and \cite{Kolo00}). The effect of the
externally applied force is simply to change the overall tilt in
the potential.

For a motor on a homogeneous or periodic substrate the effective
potential generated by the coupling to the chemical potential is
thus qualitatively the same as that of a polymer translocating
through a pore. Again on long time scales and large length scales
the dynamics is just diffusion with drift. The equation for the
velocity and diffusion constant are given by Eqs.
\ref{eq:vel},\ref{eq:diff1},\ref{eq:diff2} together with the rates
displayed in Eq. \ref{eq:rates}.

As for the polymer translocation problem, it is interesting to
consider various limits for the case $f=0$. Using (\ref{eq:rates})
we find in the limit of $\Delta \mu /T \ll 1$ a drift velocity
\begin{equation}
v=\frac{2 a_0 (\omega' \alpha-\omega \alpha')e^{-\Delta
\varepsilon /T}}{(\alpha+\omega+\alpha'+\omega')(1+e^{-\Delta
\varepsilon /T})} \left( \frac{\Delta \mu}{T} \right)
\end{equation}
Therefore, for small $\Delta \mu /T$, the velocity decreases as
$\Delta \varepsilon$ increases. Note that even when $\Delta \mu
\neq 0$, $v$ vanishes for a symmetric substrate, i.e. for
$\omega'=\omega$ and $\alpha'=\alpha$. A natural measure of the
asymmetry of the potential is $\omega' \alpha / \omega \alpha'$.
When this quantity is greater than one (less than one) a positive
$\Delta \mu$ induces a motion to the right (left). This result
remains valid to any order in $\Delta \mu$.

The maximum possible motor velocity $v_{\rm max}$ is obtained in
the limit $\Delta \mu /T \to \infty$ and $\Delta \mu \gg \Delta
\varepsilon$, where
\begin{equation} v_{\rm max}=2 a_0 \frac{\omega'
\alpha-\omega \alpha'}{\alpha+\alpha'} \;.\label{eq:vmax}
\end{equation}
In contrast to the previous regime and the polymer translocation
problem, the velocity is insensitive to $\Delta \varepsilon$.
Because of the injection of large amounts of external chemical
energy, the barrier $\Delta \varepsilon$ no longer controls a rate
limiting step.

For the diffusion constant of this model of molecular motors in
the limit $\Delta \mu / T \ll 1$ we find
\begin{eqnarray}
D&=&4 a_0^2 \frac{(\alpha+\omega)(\alpha'+\omega')e^{-\Delta
\varepsilon
/T} }{(\alpha+\omega+\alpha'+\omega')(1+e^{-\Delta \varepsilon /T})} \nonumber \\
&+& a_0^2 \frac{\Delta \mu}{T}\frac{2 e^{-\Delta \varepsilon
/T}G}{(\alpha+\omega+\alpha'+\omega')^2(1+e^{-\Delta \varepsilon
/T})^2}
\end{eqnarray}
with
\begin{eqnarray}
G &=& e^{-\Delta \varepsilon
/T}(\alpha+\omega-\alpha'-\omega')(\alpha'\omega-\alpha\omega')
\nonumber \\
&+& \alpha\alpha'(2(\alpha+\alpha')+3(\omega+\omega'))
\nonumber \\
&+&(\omega+\omega')(\alpha'\omega+\alpha\omega')+
\alpha'^2\omega+\alpha^2\omega'
\end{eqnarray}
Like the velocity, the diffusion constant decreases as $\Delta
\varepsilon$ increases in this regime. Note that the diffusion
constant increases as $\Delta \mu$ increases, because $\Delta \mu$
enhances the rates of motion in both directions. In the limit
$\Delta \mu /T \to \infty$ one obtains
\begin{equation}
D_{\rm max}=2 a_0^2 \frac{ \omega \alpha' + \omega' \alpha+
2\alpha \alpha' }{\alpha+\alpha'} \;.
\end{equation}
Again, for large chemical potential differences the result is
independent of $\Delta \varepsilon$.

\section{The Effect of Heterogeneity on the Energy Landscape}

Next we discuss the effect of heterogeneity on the effective
energy landscape experienced by motors or translocating polymers.
The detailed dynamics which results will be considered in Sec. IV.
As we shall see, heterogeneity has dramatic consequences over a
range of parameters close to the stall force.

We first consider the somewhat simpler problem of heterogeneity
and polymer translocation. We then show that a similar picture
arises for motor proteins on heterogeneous substrates like DNA or
RNA.

\subsection{Polymer Translocation}
\label{translocation-rand}

Two sources of heterogeneity affect polymer translocation. Both
arise for polymers  composed of different types of monomer. We
assume for simplicity that the monomers composing the polymer are
drawn from some random distribution with a finite variance.
Provided the correlations along the backbone are short range our
results are insensitive to the exact nature of the distribution.
The effect of sequence heterogeneity corresponding to a {\it
particular} nucleotide sequence could easily be incorporated into
a numerical analysis of the dynamics.

\begin{figure}
\includegraphics[width=8cm]{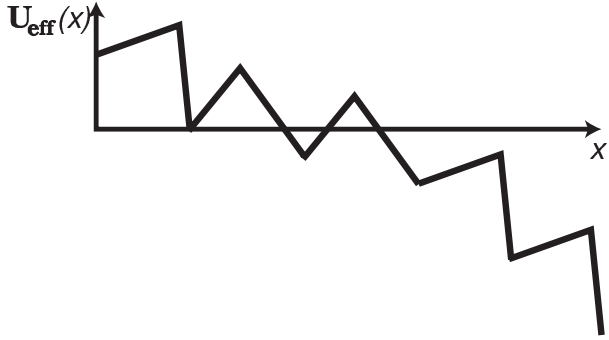} \caption{Graphical representation
of the energy landscape in the case of heterogeneous polymer
translocation when the chemical environments on both sides of the
pore are different. Potential fluctuations about the mean slope
scale like $\sqrt{x}$ for large $x$. The same picture holds for
molecular motors moving on a heterogeneous substrate powered by a
finite chemical potential difference. \label{fig:polypotran} }
\end{figure}

We first consider general features of the potential for a model
with sequence heterogeneity. Randomness in the composition of the
polymer will of course modify the interaction potential between
the polymer and pore, $U(x)$. It is easy to see that this leads to
a random potential component with a finite variance around its
mean value, i.e., a random energy landscape. The second, more
striking effect, arises from the randomness in the binding energy
of the proteins. The associated force depends specifically on the
location $x$ along the polymer. In a convenient continuum
notation, the total energy gained by attaching to the monomers has
the form $\int_0^x F_\mu(x') dx'$, where $F_\mu(x)$ represents the
different binding energies associated with the sequence of the
polymer. If the sequence is random the fluctuations around the
mean slope of the potential grow like $\sqrt{x}$. The effective
potential experienced by the random walker is therefore
\begin{equation}
{\bf U}_{\rm eff}(x)=U(x)-(\int_0^x F_\mu(x') dx' - Fx) \;,
\end{equation}
where we have included the externally applied force, $F$. A
schematic representation of the potential is shown in Fig.
\ref{fig:polypotran}. Since $\int_0^x F_\mu(x') dx'$ has
fluctuations which grow as $\sqrt{x}$, the sequential binding of
proteins to a translocating polymer creates a {\it random forcing}
landscape, in contrast to the landscape defined by Eq.
\ref{eq:pot}. Because the energy landscape itself can be viewed as
a simple random walk about a linear landscape, the {\em random
force} contribution to ${\bf U}_{\rm eff}(x)$ (an integrated
random walk) dominates the {\em random energy} term arising from
interactions with the pore. As will be discussed in Section IV
this results in unusual behavior if the externally applied force
lies in a certain range of values near the stall force.

Note that it is also possible to obtain a purely random energy
landscape in polymer translocation. When the chemical environments
on both sides match (e.g., for identical concentrations of binding
proteins) one has $F_\mu(x)=0$. The only random component of the
energy landscape is due to the potential for translocating through
the pore which has bounded fluctuations about its mean value. For
this energy landscape, the dynamics at long times and large length
scales is then biased diffusion, with a drift velocity and
diffusion constant renormalized by the heterogeneous interactions
with the pore \cite{Alex81}.

We now explore these effects within our microscopic model of
polymer translocation. The heterogeneity is introduced into the
model through the rates (\ref{eq:polyrates}). Imagine drawing the
set of parameters $\{p\}=\{\omega,\omega',\Delta \varepsilon,
\Delta \mu \}$ from random distributions (corresponding to various
nucleotide sequences) with a finite variance. According to Eq.
(\ref{eq:dEpol}),, the total change in energy after $m$ monomers
translocate is given by
\begin{equation}
E(m)=2fm+\sum_{l=1}^m \Delta E(l) \;.
\end{equation}
Here the $\Delta E(m)$ are effective energy differences between
two even sites corresponding to the set of values of the set
$\{p\}$ drawn randomly. Since the energy is a sum of independent
random variables a {\it random forcing} landscape is developed.

We expect that a simple {\it random energy} landscape results if
we turn off the protein binding by setting $\Delta \mu =0$.
However, because the energy at even sites is always $E=0$ in our
simple model, the landscape is just a uniform tilt in this limit.
A more realistic model would allow additional energy variations at
these sites. If we assign an energy $\varepsilon(m)$ to these even
sites, it is straightforward to show that the total change in
energy after $m$ monomers have translocated takes the form
\begin{equation}
E(m)=2fm+\varepsilon(m) \;,
\end{equation}
corresponding to a random energy landscape.

\subsection{Molecular Motors}

We now turn to the effect of heterogeneity on molecular motors.
Here, as for polymer translocation, we select the set of
parameters $\{p\}=\{\alpha,\alpha',\omega,\omega',\Delta
\varepsilon\}$ from a random distribution with a finite variance.
For some motors and enzymes (for example RNA Polymerase,
helicases, DNA polymerases and exonucleases -- see introduction
and below), $\Delta \mu$ may also be random. This clearly only
adds an additional contribution to the random forcing landscape.
Using the results presented above it is easy to see from Eq.
(\ref{eq:dE}) that the total effective energy change after $m$
monomers is given by
\begin{equation}
E(m)=2fm+\sum_{l=1}^m \Delta E(l)\;. \label{eq:menergy}
\end{equation}
Here, each $\Delta E(m)$ corresponds to an independent set of
values of $\{p\}$ drawn randomly. Thus, as in the polymer
translocation problem, the potential is random forcing.

For motors such as helicases, DNA polymerase and exonucleases, and
RNA polymerase and ribosomes an additional contribution to the
random energy arises due to the force associated with, e.g., base
pairing energies or the trailing strand which is produced. The
effect of this would be to modify the expression above to
\begin{equation}
E(m)=2fm+\sum_{l=1}^mf_\mu(l)+\sum_{l=1}^m \Delta E(l)\;,
\label{eq:mmenergy}
\end{equation}
where $f_\mu$ is the additional contribution of the explicit
random forcing from monomer $m$. The resulting random forcing
landscape is even more pronounced.

The above scenario applies as long as the chemical potential
difference $\Delta \mu \neq 0$. In the case when $\Delta \mu = 0$
it is easy to see that $\Delta E(m)=0$ unless we allow, as in the
polymer translocation problem, for the energy at even sites also
to vary and take the value $\varepsilon(m)$. In this case we
obtain
\begin{equation}
E(m)=2fm+\varepsilon(m) \;,
\end{equation}
corresponding to a random energy landscape provided
$\varepsilon(m)$ has only short range correlations. Although we
could write the energy in the form of Eq. \ref{eq:menergy}, now
$\Delta E(m)=\varepsilon(m)-\varepsilon(m-1)$, so $\Delta E(m)$ is
effectively the gradient of a random potential with bounded
fluctuations. Note, however, that for motors with an $f_\mu$
contribution (as in Eq. (\ref{eq:mmenergy})) it is not possible to
obtain a random energy landscape.

The energy landscape for both polymer translocation and molecular
motors is therefore qualitatively identical. Generically, in both
cases, a random forcing energy landscape develops. However, if the
motor model without the applied external force has no bias (i.e.,
if $\Delta \mu = 0$), we recover the diffusion with drift dynamics
associated with a random energy potential.

\section{Dynamics In Heterogeneous Environments}
\label{dynamics}

In this section we discuss in detail the dynamics of translocating
polymers and motor proteins with heterogeneity for the model
depicted schematically in Fig. \ref{fig:motor}. We describe four
distinct cases with different dynamical behaviors as the
externally applied force is varied. The critical forces for the
transition between the regimes can be calculated exactly in terms
of the rates
$w_a^\rightarrow,w_b^\leftarrow,w_a^\leftarrow,w_b^\rightarrow$
averaged over their heterogeneous generalization with $f=0$. The
explicit expressions for polymer translocation and molecular
motors can be easily obtained by using the rates in Eqs.
(\ref{eq:polyrates}) and (\ref{eq:rates}) respectively. We assume
throughout that $\Delta \mu \neq 0$, as the case $\Delta \mu =0$
leads only to a random energy model and biased diffusion. Also,
contributions to the random forcing energy landscape of the form
of Eq. (\ref{eq:mmenergy}) are omitted for simplicity. Their
addition is straightforward and can be easily seen to enhance the
region of anomalous dynamics.

The dynamical behaviors of random walkers in random forcing or
random energy landscapes have been studied in detail in the
statistical mechanics literature \cite{Derrida83,Bouchaud90}.
Unusual dynamical behavior arises for random walkers in a random
forcing energy landscape. Using the results of Derrida
\cite{Derrida83}, one can calculate the transition points between
the different regimes including the effect of randomness. Parts of
the calculation are outlined in Appendix \ref{random} along with
the different regimes in terms of the transition rates
$w_a^\rightarrow,w_b^\leftarrow,w_a^\leftarrow,w_b^\rightarrow$.
Here we consider the experimental setup in Figs. \ref{fig:polymer}
and \ref{fig:motor} where the external force is varied. Denoting
spatial averages by an overline and using the results of the
Appendix C one finds the following regimes.

\noindent {\bf Regime I}: The velocity $v$ and diffusion constant
$D$ of the model are finite when
\begin{equation}
f<-\frac{T}{4} \ln \overline{ \left( \frac{w_a^\leftarrow
w_b^\leftarrow}{w_a^\rightarrow w_b^\rightarrow} \right)^2 }_{f=0}
\;,
\end{equation}
or
\begin{equation}
f>\frac{T}{4} \ln \overline{ \left( \frac{w_a^\rightarrow
w_b^\rightarrow}{w_a^\leftarrow w_b^\leftarrow} \right)^2 }_{f=0}
\;,
\end{equation}
where the subscript $f=0$ denotes that $f$ has been set to zero in
the average. In this regime $ \langle x \rangle  =vt$ and $\langle
x^2\rangle - \langle x \rangle^2 =2Dt$ for long times, where the
angular brackets denote an average over different thermal
histories of the system. Simpler conditions can be obtained by
assuming that $\Delta E(m)=T\ln\left((w_a^\leftarrow
w_b^\leftarrow)/(w_a^\rightarrow w_b^\rightarrow)\right)$ has a
{\it Gaussian} distribution about the mean $2f+\overline{\Delta
E}_{f=0}$ (see Eqs. \ref{eq:dEpol} and \ref{eq:dE}) and a variance
$V= \overline{(\Delta E)^2}_{f=0} - \overline{(\Delta
E)}_{f=0}^2$. Here again the subscript $f=0$ denotes that averages
are taken with the value of the force set to zero. In the case one
has
\begin{eqnarray}
f&>& \frac{1}{2}\left(\overline{\Delta E}_{f=0}+V/T \right)\;,
\nonumber \\
f&<& \frac{1}{2}\left(\overline{\Delta E}_{f=0}-V/T \right)\;.
\end{eqnarray}
Note that the force does not contribute to the variance so that
$V=\overline{\Delta E^2}_{f=0}-\overline{\Delta
E}^2_{f=0}=\overline{\Delta E_f^2}-\overline{\Delta E_f}^2$.

\noindent {\bf Regime II}: The velocity $v$ is finite but the
diffusion constant is infinite. Thus, in this region $ \langle x
\rangle =vt$ and $\langle x^2 \rangle - \langle x\rangle^2 \sim
t^{2/\mu}$, where $1<\mu<2$. The relevant force ranges are
\begin{equation}
 -\frac{T}{4} \ln \overline{ \left( \frac{w_a^\leftarrow
w_b^\leftarrow}{w_a^\rightarrow w_b^\rightarrow} \right)^2 }_{f=0}
< f \leq -\frac{T}{2} \ln \overline{ \left( \frac{w_a^\leftarrow
w_b^\leftarrow}{w_a^\rightarrow w_b^\rightarrow} \right) }_{f=0}
\;,
\end{equation}
and
\begin{equation}
\frac{T}{2} \ln \overline{ \left( \frac{w_a^\rightarrow
w_b^\rightarrow}{w_a^\leftarrow w_b^\leftarrow} \right) }_{f=0}
\leq f <  \frac{T}{4} \ln \overline{\left( \frac{w_a^\rightarrow
w_b^\rightarrow}{w_a^\leftarrow w_b^\leftarrow} \right)^2 }_{f=0}
\;.
\end{equation}
Provided that $\Delta E$ has a Gaussian distribution the
conditions reduce to
\begin{eqnarray}
\frac{1}{2}\left(\overline{\Delta E}_{f=0}+V/2T \right)&<&f\leq
\frac{1}{2}\left(\overline{\Delta E}_{f=0}+V/T \right)\;,
\nonumber
\\
\frac{1}{2}\left(\overline{\Delta E}_{f=0}-V/T \right)&\leq&f<
\frac{1}{2}\left(\overline{\Delta E}_{f=0}-V/2T \right)\;.
\end{eqnarray}
For a Gaussian distribution it is known \cite{Bouchaud90} that the
exponent $\mu$ is given by
\begin{equation}
\mu=2T\vert \overline{\Delta E}_{f=0}-2f\vert/V.
\end{equation}

\noindent {\bf Regime III}: The velocity $v$ is zero in the sense
that $ \langle x \rangle \sim t^\mu$, where $\mu<1$. The exponent
$\mu$ also controls the variance, $\langle x^2 \rangle - \langle x
\rangle^2 \sim t^{2 \mu}$. This behavior occurs when
\begin{equation}
-\frac{T}{2} \ln \overline{\left( \frac{w_a^\leftarrow
w_b^\leftarrow}{w_a^\rightarrow w_b^\rightarrow} \right) }_{f=0}
\leq f \leq \frac{T}{2} \ln \overline{ \left(
\frac{w_a^\rightarrow w_b^\rightarrow}{w_a^\leftarrow
w_b^\leftarrow} \right)}_{f=0} \;.
\end{equation}
When $\Delta E$ has a Gaussian distribution, these conditions
reduce to
\begin{equation}
\frac{1}{2}\left(\overline{\Delta E}_{f=0}-V/2T
\right)<f<\frac{1}{2}\left(\overline{\Delta E}_{f=0}+V/2T
\right)\;. \label{eq:frange}
\end{equation}

\noindent {\bf Sinai diffusion}: Here $\langle x \rangle=0$ and
$\langle x^2 \rangle \sim (\ln(t/\tau))^4$, where $\tau$ is the
microscopic time needed to move across one monomer.  This regime
appears precisely at the ``stall force'' corresponding to a
disordered substrate, namely
\begin{equation}
f_s = \frac{T}{2} \overline{ \ln  \left( \frac{w_a^\rightarrow
w_b^\rightarrow}{w_a^\leftarrow w_b^\leftarrow} \right) }_{f=0}
\;.\label{eq:sinailoc}
\end{equation}
If $\Delta E$ has a Gaussian distribution this condition yields
\begin{equation}
f_s=\frac{\overline{\Delta E}_{f=0}}{2} \;. \label{eq:stall}
\end{equation}
The resulting behavior as the force is varied is summarized
qualitatively in Fig. \ref{fig:Phasedia}.

It is interesting to consider the location of the stall force,
$f_s$, as well as the range of forces over which the displacement
is anomalous, namely the region where $v=\lim_{t \to \infty}
\langle x \rangle / t=0$, in some more detail for both polymer
translocation and molecular motors in some simple scenarios. These
quantities characterize how the location and width of the
anomalous displacement region develops as a function of
temperature and chemical forces. We assume $\Delta E(m)$ with a
Gaussian distribution about $\overline{\Delta E}$ with a variance
$V$, although it is straight forward to extend the results to
non-Gaussian distributions with no change of the qualitative
behavior. It is straightforward to show using Eq.
(\ref{eq:frange}) that the range of forces, $\Delta f$, over which
the velocity is zero satisfies
\begin{equation}
\Delta f =\frac{1}{2T}~V \;. \label{eq:frangeest}
\end{equation}

For polymer translocation using Eqs. (\ref{eq:dEpol}) and
\ref{eq:stall} implies that Sinai diffusion occurs for the force
\begin{equation}
f_s=\frac{\overline{\Delta \mu}}{2} \;,
\end{equation}
while Eq. \ref{eq:frangeest} implies that the range of forces
around $f_s$ where the displacement is anomalous is given by
\begin{equation}
\Delta f =\frac{1}{2T}\left(\overline{\Delta
\mu^2}-\overline{\Delta \mu}^2 \right)\;.
\end{equation}
If there are no proteins on lefthand side (cis chamber) and a
small concentration, $P$, of protein is added to the righthand
side (trans chamber) one can show using (\ref{eq:polchempot}) that
$f_s \propto P$ while $\Delta f \propto P^2$. Thus, as the
chemical bias increases both $f_s$ and $\Delta f$ grow. Note that
in general one may consider proteins in both the left and right
chambers. In this case even when the average chemical bias
$\overline{\Delta \mu}=0$ one may still have $V >0$ giving rise to
anomalous dynamics even when the external bias $F=0$.

For molecular motors the situation is more interesting. The
results presented above  for the transition points between the
different regimes hold even when $\Delta \mu$ is also random.
However, here we restrict ourselves to the simpler case when
$\Delta \mu$ is constant. In this case (\ref{eq:dE}) implies that
for small chemical potential ($\Delta \mu/ T \ll 1$) the chemical
energy difference $\Delta E_{f=0} = q \Delta \mu$, where $q$ is
the coefficient in the Taylor expansion of (\ref{eq:dE}) in
$\Delta \mu$ which is independent of $T$. Therefore, in this
limit, the stall force is
\begin{equation}
f_s=\frac{\overline{q} \Delta \mu}{2} \;,
\end{equation}
and
\begin{equation}
\Delta f =\frac{\Delta
\mu^2}{T}\left(\overline{q^2}-\overline{q}^2 \right)\;,
\end{equation}
where we have assumed the purpose of a rough estimate that the
chemical potential difference does not depend on the type of
monomer. Similarly to polymer translocation, as the system is
driven out of chemical equilibrium both $f_s$ and $\Delta f$ grow.
However, in the limit of $\Delta \mu/T \gg 1$ one obtains $\Delta
E_{f=0} = p T$ where $p$ is obtained by taking the appropriate
limit in (\ref{eq:dE}) and is independent of $T$. We then have
\begin{equation}
f_s=\frac{\overline{p} T}{2} \;,
\end{equation}
and
\begin{equation}
\Delta f =T\left(\overline{p^2}-\overline{p}^2 \right)\;,
\end{equation}
implying that both quantities {\it increase} with increasing
temperature.

Note that if the force applied to the polymer or motor is held
constant and the chemical parameters (e.g. ATP or protein
concentration) are varied from their equilibrium value one should
also observe a region of anomalous displacement (see the general
expressions in Appendix \ref{random}). These conclusions are
summarized qualitatively in Figs. \ref{fig:Phasedia},
\ref{fig:changechem} and \ref{fig:BigDia}.

\section{Experimental Considerations}

As discussed in the previous section, the important quantity for
deciding if anomalous dynamics is present is the variance $V
\equiv \overline{(\Delta E^2)}  - \overline{(\Delta E )}^2$ of
$\Delta E(m)$, where the overbar represents an average over the
ensemble of random sequences. Effects related to sequence
heterogeneity dominate when $V$ is large compared to $k_BT
\overline{ \Delta E}$. Here we estimate the ranges over which
anomalous dynamics may be observed in experiments as well as other
preconditions needed to observe this behavior. We also discuss the
effect of finite time experiments on the shape of the
velocity-force curve. In this section, we reintroduce Boltzmann's
constant $k_B$.

\subsection{Polymer Translocation}
For polymer translocation, whether the variance $V$ is large
compared to $k_BT \overline{ \Delta E }$ of course depends on a
number of factors, including the base composition of the
polynucleotide passing through the pore, the particular protein
whose binding drives translocation, and the concentration of the
binding protein. Nonetheless, it is instructive to consider an
example to get some sense of the orders of magnitude involved. We
focus on DNA binding proteins. Note that, like those of most such
proteins, the binding sites are several nucleotides long; unlike
in previous sections, unless stated otherwise, we will give values
of $V$ and other parameters normalized per nucleotide rather than
per bound protein.

The bacteriophage T4-coded gene 32 protein (gp32) is a monomeric
single-stranded DNA (ssDNA) binding protein which is implicated in
DNA replication and related processes~\cite{Coleman80}.  When it
associates with ssDNA cooperatively in the ``polynucleotide''
mode~\cite{Kowal81}, its net affinity~\footnote{The net affinity
is the affinity of an additional protein molecule for a growing
chain of cooperatively bound monomers; it differs from the
affinity of an isolated protein molecule for ssDNA by an
enhancement factor arising from the cooperative interactions}
\knet\ can vary by as much as a factor of 10 depending on the
polymer's base composition; in physiological salt concentrations,
a typical range is $\knet \sim 2 \times 10^8$ -- $2 \times 10^9
~\mathrm{M}^{-1}$~\cite{Newport81}.  In this binding mode, the
binding site of each gp32 monomer is 7 nucleotides long. For a
$\mu \mathrm{M}$ protein concentration, \knet\ is large enough
that almost all sites on the translocated ssDNA will be occupied.
Upon assuming that $V$ is determined entirely by the base
dependence of \knet, we then estimate that $V \sim 0.1$ -- $0.2
(k_BT)^2$ for a ``generic'' DNA molecule in which each of the
bases appears with roughly equal frequency. Here $T$ is room
temperature, $k_B T\simeq 0.59 kcal/mole$. In this case, the
change in free energy of a nucleotide moved from a buffer without
any gp32 to one where the protein is present is $\Delta \mu \sim
k_BT$ (see Eq.~\ref{pore-omegas}). Upon taking the ssDNA to be a
freely-jointed chain with Kuhn length 1.5 nm~\cite{Smith96}, one
finds that a force of about 10--15 pN on the polymer is required
to cancel the effects of the protein binding.  In order to have
$k_BT \overline{\Delta E} \lesssim V$, so that disorder effects
can be detected, the value of the force must be controlled to
roughly 10\% or better accuracy.

\subsection{Molecular Motors}
To be able to measure the motion of a motor along a substrate it
must remain attached long enough to be able to preform many moves
across monomers. In other words, if the rate at which the motor
leaves the substrate is $\gamma$ and the rate of crossing a
monomer to the right or left is $w^\rightarrow$ or $w^\leftarrow$
respectively, then $\gamma \ll w^\rightarrow + w^\leftarrow$ must
hold. In the regime of anomalous dynamics $w^\rightarrow$ is of
the same order of $w^\leftarrow$. Therefore, the condition will
not be fulfilled in this regime when the rate of hopping against
the chemical bias $w^\leftarrow$ is always very small.

There are, however, experiments where such a restriction does not
hold. For example, the experiment by Wuite et al. \cite{Gijs00} on
the DNA polymerase/exonuclease system (see also \cite{Maier00})
monitors not the displacement of a single motor but the location
of the {\it junction} between the ssDNA and the dsDNA. Therefore,
it is more natural to model the dynamics of the ssDNA/dsDNA
junction and not of the motor. A motor which leaves the
ssDNA/dsDNA junction is eventually replaced by a motor from the
solution. Within our models this can be represented by an internal
state of the junction (similar in spirit to \cite{Fisher99} and
\cite{Kolo00}). A model of this type for the DNA
polymerase/exonuclease has been studied in \cite{Anita03}.
However, the disorder in the transition rates, present due to the
heterogeneity of the DNA, has been neglected. In Appendix D we
analyze in some detail a simple model of the DNA
polymerase/exonuclease system. As shown in the appendix it is
straight forward to show that the presence of heterogeneity (for
example, in the energy gained from the hydrolysis of the different
NTP's) leads to a random forcing energy landscape. One therefore
expects a region of anomalous dynamics near where the external
stretching force $F'$ causes a change in direction. We stress that
more realistic models with many intermediate states can by
analyzed similarly without affecting the existence of the region
with anomalous dynamics. Unfortunately, for this experiment an
estimation of the width of the region is not straightforward.

Estimates, similar to those above for polymer translocation, can
be obtained for the random force landscapes for a number of
molecular motors which operate on DNA or RNA. Two examples of
interest are RNA polymerases
(RNAp's)~\cite{Wang98,Julicher98,Gelles98,Davenport00} and
helicases acting on double-stranded DNA
(dsDNA)~\cite{Lohman96,vonHippel01,Bianco01,Dohoney01}.  An RNAp's
function is to transcribe DNA---that is, to synthesize an RNA
``copy'' with the same sequence as a DNA molecule.  To do so, it
walks along dsDNA trailing a growing RNA strand.  The RNAp motor
is powered entirely by the energy gained from the hydrolysis of
successive nucleotide triphosphates (NTP's) as they are added to
the RNA molecule.  Although the mechanism of RNAp motion is still
the subject of debate~\cite{Julicher98,vonHippel02}, many models
suggest that at low enough NTP concentrations, its ability to move
forward will be limited by the rate at which NTP's arrive at the
catalytic site.  A straightforward way to force RNAp into a regime
in which its motion is dominated by a random force energy
landscape is thus to place it in a buffer with different
concentrations of each of the four NTP's. The motor's ability to
take a forward step is dependent on the incorporation of the
appropriate NTP, and the rate of that incorporation is
proportional to that NTP's concentration. Thus, one can in
principle make $V$ arbitrarily large and satisfy the criterion
$V>k_bT \overline{\Delta E}$ for significant random force effects.
Each factor of 10 difference between the concentrations of two
nucleotide triphosphates, and hence in the rates to make a forward
step, translates into a difference of $k_B T \ln(10) \approx 2.3
k_BT$ in $\Delta E(m)$. Of course, in practice other factors---for
example the possibility that the RNAp might fall off its DNA track
before the needed NTP arrives---will limit how large a range of
concentration differences can be achieved experimentally. It will
be interesting to see whether strong disorder effects can be
observed.

Another class of motors that use DNA as their track are helicases,
which are needed to separate the two strands of dsDNA in order to
facilitate various processes in the cell such as cell division in
prokaryotes. Helicases move along the DNA by consuming energy from
NTP's. While some helicases only break a few base pairs at a time,
others can move substantial distances along their
tracks~\cite{Bianco01,Dohoney01}.  Recent modelling of certain
monomeric helicases~\cite{Betterton02} suggests that disordered
DNA sequences affect helicase motion primarily through the
different energies required to open different base pairs.  Random
sequences thus lead to anomalous helicase motion in much the same
way they do anomalous dynamics of mechanical
unzipping~\cite{Lubensky00,Lubensky02}.  In the simplest case of
``passive'' opening, one finds that $\Delta E(m) \approx \demot +
\dedna(m)$, where $\demot$, which summarizes the forward force
exerted by the helicase motor, is negative and has magnitude at
least $\sim 2 k_BT$, and $\dedna$ is simply the thermodynamic free
energy cost of opening each successive base pair, with size
roughly between 1 and 3 $k T$~\cite{SantaLucia98}. One thus has $V
\sim 1 (k_BT)^2$.  This large variance means that it should be
relatively easy to observe anomalous, disorder-dominated dynamics
in helicases as predicted earlier for DNA unzipping.  If, for
example, one assumes that the magnitude of $\demot$ is near its
lower bound of $2 k_BT$, then, in the passive opening model,
disorder effects should begin to appear for a mechanical load
opposing the motor's motion of as little as 7 pN and should
persist up to at least 20 pN.

\subsection{Finite time effects}
\begin{figure}
\includegraphics[width=8cm]{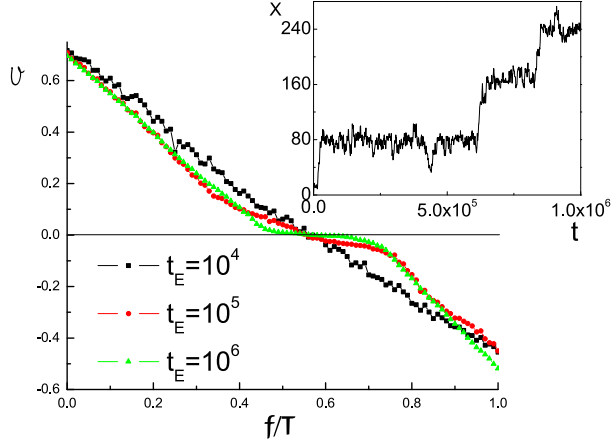} \caption{The velocity as a
function of $f/T$ for different values of $t_E$. Here $\Delta \mu
/T=3$ and parameters where chosen with equal probability to be
either $\{p\}=\{5,1,0.3,1,0\}$ or $\{p\}=\{4,0.1,0.7,1,0\}$ (see
text for notation). The calculated regime of anomalous velocity is
$0.5116<f<0.699$. Data was averaged over a 100 thermal
realizations. Inset: a single trajectory shown for the same
parameters at $f/T=0.45$. \label{fig:simu}}
\end{figure}

All calculations of quantities such as the velocity have been done
by taking the limit of very large times and averaging over thermal
realizations with the same heterogeneous sequence. For experiments
done over finite times the velocity will not be strictly zero in
the regime of anomalous dynamics. Instead, the velocity decays to
zero as $t_E^{\mu-1}$, where $t_E$ is the experimental averaging
time used to define $v$ as $\langle x(t_E)-x(0) \rangle / t_E$.
The closer $\mu$ is to zero, the faster the decay will be.
Therefore, the curve of the velocity as a function of the external
force or chemical potential (see Figs. \ref{fig:Phasedia} and
\ref{fig:changechem}) will be rounded becoming sharper and sharper
as $t_E \to \infty$.
To illustrate this we have carried out simulations of model
(\ref{eq:rates}) on a single realization of the disorder averaging
over thermal realizations and measured the $v-F$ curve. The
results are shown in Fig. \ref{fig:simu}. As can be seen the
longer $t_E$ the closer is the $v-F$ curve to that shown in Fig.
\ref{fig:Phasedia}. The convex shape of the curve near the stall
force is clear already for averaging times $t_E \sim 10^5$,
corresponding to motors which transverse distances of $O(1000)$ at
$f/T=0$. Note that, if one looks at the displacement of a single
motor (i.e., without averaging over thermal realizations), the
regime of anomalous dynamics will be characterized by long pauses
at localized regions (corresponding to deep minima of the
effective potential) with fast transitions between the localized
regions (corresponding to overcoming the barrier associated with
the minima). The inset of Fig. \ref{fig:simu} shows a single
trajectory of as a function of time for a given realization of
disorder. The value of $f/T$ was chosen to be in the region close
to the anomalous velocity regime but not inside it (the point is
at the edge of the anomalous diffusion region close to the normal
diffusion region). As can be seen the motion of the motor is
characterized by long pauses at specific locations along the
track, with quick jumps between the pause points. The location of
the pause points is reproducible for the same spatial disorder and
different thermal realizations although their duration varies from
simulation to simulation. Note, that since the velocity is finite
in this regimes, over large length scales the effect of the jumps
becomes unimportant. These pauses correspond to local minima of
the effective potential and as such are inherently correlated with
the structure of the track. Such pauses and jumps have been
observed in recent experiments \cite{Claudia} on DNA unzipping.

\section{Summary}
We have studied the effect of sequence heterogeneity on both
polymer translocation and the motion of molecular motors within
simple models. The models were solved exactly both with and
without disorder. It was shown that these systems can be
represented on large length scales and long times by a random
walker moving along a random forcing energy landscape. Thus, in a
range of forces near the stall force we expect anomalous dynamics
where the displacement grows as a sublinear power of time. We
stress again that such results also apply to more sophisticated
models which include many internal states of the motor (see the
discussion of the DNA polymerase / exonuclease system in Appendix
D). Several systems in which the regime of anomalous dynamics
might be wide enough to be observable were considered.

{\bf Acknowledgments:} It is a pleasure to acknowledge helpful
conversations with S. Block, A. Meller and S. Xie. We are grateful
to A. Goel for a copy of \cite{Anita03} prior to publication and
to G. Lattanzi and A. Maritan for pointing out a misprint. Work by
Y.K and D.R.N was supported by the National Science Foundation
through Grant DMR-0231631 and the Harvard Materials Research
Laboratory via Grant DMR-0213805. Y.K also thanks the Fulbright
program in Israel for financial support.

\appendix
\section{The chemical potential difference for translocating polymers}
\label{chempot}

Here we discuss the dependence of the chemical potential
difference $\Delta \mu$ for a translocating polymer between the
righthand (trans) and lefthand (cis) sides of Fig.
\ref{fig:polymer} on the protein concentrations and its binding
energy to the polymer. Consider first a denatured polymer in a
solution with a concentration $c_p$ of proteins which can bind to
its monomers with a binding energy $E_b<0$. We neglect
cooperativity in the binding of the proteins to the polymer,
although this effect could easily be included. Assuming an ideal
solution theory the protein chemical potential is given
$\mu=\mu_0+T \ln \left( P \right)$, where $P=c_p/c$ and $c$ is the
concentration of the solvent. Here we take the free-energy change
due to an addition of one isolated protein to the solvent to be
$\mu_0-T\ln n$ where $n$ is the number of solvent molecules
\cite{LL}. Next, we take the energy function of a polymer of
length $N$ inside the solution to be
\begin{equation}
H=\sum_{i=1}^{N} \left(-E_b\sigma_i+\mu' \sigma_i \right)\;,
\end{equation}
where $\sigma_i=1 (0)$ if a protein is bound (unbound) to monomer
$i$ and $\mu'$ is a chemical potential which controls the density
of proteins bound to the polymer. In thermal equilibrium
$\mu=\mu'$ which gives for the free energy of a polymer monomer in
the solution
\begin{equation}
-T \ln \left(1+P \exp(\mu_0-E_b)/T) \right) \;.
\end{equation}
The change in the free energy of the polymer which occurs as a
result of a monomer passing from the left (cis) chamber to the
right (trans) chamber, with ratios of protein to solvent
concentrations $P_L$ and $P_R$ respectively, is given by
\begin{equation}
\Delta \mu =\frac{d{\cal F}}{dN_R}= -T \ln \left( \frac{1+P_L
\exp(\mu_0-E_b)/T)}{1+P_R \exp(\mu_0-E_b)/T)} \right) \;.
\label{eq:polchempot}
\end{equation}
where ${\cal F}$ is the total free energy of the polymer and $N_R$
is the number of monomers in the right chamber. It is
straightforward to see that this result implies that for proteins
with different binding energy to different types of monomers
$\Delta \mu$ will depend on the type of monomer.

\section{Deriving the effective potential from the master equation}
\label{master}

In this appendix we show that the equations for the probability
$P_n(t)$ of being at site $n$ at time $t$ are equivalent in the
long-time limit (to be specified more exactly below) to a random
walker moving in an energy landscape constructed using Eq.
(\ref{eq:dEdef}). We demonstrate this by eliminating the even
sites from the equations of motion (see \cite{Lattanzi02} for
similar ideas).

First, consider the equations governing the evolution of the
probability, i.e. the master equation. For odd $n$ one has (see
Fig. \ref{fig:motor})
\begin{equation}
\frac{dP_n(t)}{dt}= w_a^\rightarrow P_{n-1}(t)+ w_a^\leftarrow
P_{n+1}(t) - (w_b^\rightarrow+w_b^\leftarrow) P_{n}(t) \;,
\end{equation}
while for even $n$
\begin{equation}
\frac{dP_n(t)}{dt}= w_b^\rightarrow P_{n-1}(t)+ w_b^\leftarrow
P_{n+1}(t) - (w_a^\rightarrow+w_a^\leftarrow) P_{n}(t) \;.
\end{equation}

Next, we solve the equation for the odd sites and substitute into
that for the even sites. The solution of the equation for the odd
sites is
\begin{equation}
P_n(t)=e^{-(w_b^\rightarrow + w_b^\leftarrow)t} \left( P_n(0) +
\int_0^t d\tau e^{(w_b^\rightarrow + w_b^\leftarrow) \tau}
\left(w_a^\rightarrow P_{n-1}(\tau)+ w_a^\leftarrow
P_{n+1}(\tau)\right) \right) \;,
\end{equation}
where $P_n(0)$ is the probability distribution at the initial time
$t=0$. Substituting this into the equation for the even sites
yields
\begin{eqnarray}
\frac{dP_n(t)}{dt} &=& e^{-(w_b^\rightarrow + w_b^\leftarrow)t}
\int_0^t d\tau e^{(w_b^\rightarrow + w_b^\leftarrow) \tau} \left(
w_b^\rightarrow w_a^\rightarrow P_{n-2}(\tau) + w_b^\leftarrow
w_a^\leftarrow P_{n+2}(\tau) \right) \nonumber \\  &+&
e^{-(w_b^\rightarrow + w_b^\leftarrow)t} \int_0^t d\tau
e^{(w_b^\rightarrow + w_b^\leftarrow) \tau}\left(w_b^\rightarrow
w_a^\leftarrow+w_b^\leftarrow w_a^\rightarrow \right)P_n(\tau) \label{eq:dec} \\
&-& (w_a^\rightarrow+w_a^\leftarrow)P_n(t)+e^{-(w_b^\rightarrow +
w_b^\leftarrow)t}(w_b^\rightarrow P_{n-1}(0) + w_b^\leftarrow
P_{n+1}(0)) \nonumber \;.
\end{eqnarray}
At times $t \gg (w_b^\rightarrow + w_b^\leftarrow)$ one can
neglect the two last terms in (\ref{eq:dec}) and approximate the
integrals as follows
\begin{equation}
\int_0^t d\tau e^{(w_b^\rightarrow + w_b^\leftarrow) \tau} f(\tau)
\approx \frac{1}{(w_b^\rightarrow + w_b^\leftarrow)}
e^{(w_b^\rightarrow + w_b^\leftarrow) t} f(t)
\end{equation}
where $f(\tau)$ is assumed to vary slowly with $\tau$. In this
long time approximation, Eq. (\ref{eq:dec}) reduces to

This reduces (\ref{eq:dec}) to
\begin{equation}
\frac{dP_n(t)}{dt}=w_b^\leftarrow w_a^\leftarrow P_{n+2}(t)+
w_b^\rightarrow w_a^\rightarrow P_{n-2}(t)-(w_b^\rightarrow
w_a^\rightarrow+w_b^\leftarrow w_a^\leftarrow)P_n(t)
\end{equation}
where we have rescaled times such that $t \to
\frac{t}{(w_b^\rightarrow + w_b^\leftarrow)}$. As expected this
equation corresponds to a random walker moving in a potential
constructed using Eq. (\ref{eq:dEdef}).

\section{Derivation of the Different Dynamical Regimes}
\label{random}

In this appendix the expressions for the different dynamical
regimes in terms of the hopping rates
$w_a^\rightarrow,w_a^\leftarrow,w_b^\rightarrow,w_b^\leftarrow$
are given. These general equations allow a straightforward
derivation of the expressions in the text. However, before turning
to the results we outline the derivation of the regime where the
displacement is anomalous. The derivation of the other regimes is
much lengthier, so we only sketch the main results.

Unless stated otherwise we assume throughout this appendix that
\begin{equation}
 \overline{\log \left( \frac{w_a^\leftarrow
w_b^\leftarrow}{w_a^\rightarrow w_b^\rightarrow} \right)}
 < 0 \;,
\end{equation}
where, as in the main text, we denote spatial averages by an
overbar. Because $\exp(-\Delta E /T)=w_a^\leftarrow w_b^\leftarrow
/ w_a^\rightarrow w_b^\rightarrow$ in our notation, this condition
is equivalent to assuming an overall bias to the right
\begin{equation}
\overline{ \Delta E}  < 0 \;,
\end{equation}
where $\Delta E$ arises from the generalization of Eqs.
(\ref{eq:dEpol}) and (\ref{eq:dE}) to heterogeneous systems. The
other opposite regime $\overline{\Delta E}>0$ can be treated
similarly. As shown by Derrida \cite{Derrida83} the velocity of a
random walker on an infinite lattice model in this case is given
by
\begin{equation}
v=\lim_{N\to \infty} \frac{N}{\sum_{i=1}^{N} r_i} \;,
\label{eq:derv}
\end{equation}
where
\begin{equation}
r_i=\frac{1}{W_{i+1,i}} \left[ 1 + \sum_{k=1}^{N-1}
\prod_{l=1}^{k} \left( \frac{W_{i+l-1,i+l}}{W_{i+l+1,i+l}}\right)
\right] \;. \label{eq:derprob}
\end{equation}
Here $W_{i,j}$ is the hopping rate from site $j$ to $i$. The
denominator of (\ref{eq:derv}) can be simplified by replacing the
sum by an average of $r_i$
\begin{equation}
\langle r \rangle = \lim_{N \to \infty} \frac{1}{N} \sum_{i=1}^N
r_i \;.
\end{equation}
Using the rates
$w_a^\rightarrow,w_a^\leftarrow,w_b^\rightarrow,w_b^\leftarrow$
one finds that the average $\langle r \rangle$ is finite only if
\begin{equation}
\overline{\left( \frac{w_a^\leftarrow
w_b^\leftarrow}{w_a^\rightarrow w_b^\rightarrow} \right)} < 1 \;.
\end{equation}
In this case the velocity is finite. However, when the inequality
is reversed $\langle r \rangle = \infty$ and the velocity is zero.

A much lengthier calculation along somewhat similar lines can be
done to derive the other dynamical regimes. One obtains the
following results.

\noindent {\bf Regime I}: When
\begin{equation}
 \overline{\left( \frac{w_a^\leftarrow
w_b^\leftarrow}{w_a^\rightarrow w_b^\rightarrow} \right)^2}
 < 1 \;,
\end{equation}
the velocity $v$ and diffusion constant $D$ of the model are
finite. Namely, $ \langle x \rangle  =vt$ and $\langle x^2\rangle
- \langle x \rangle^2 =2Dt$ for long times, where the angular
brackets denote an average over different thermal histories of the
system. Assuming for simplicity that $\Delta E(m)$ is distributed
around $\overline{\Delta E}$ with a Gaussian distribution with a
variance $V= \overline{(\Delta E)^2 } - \overline{(\Delta E)}^2$
this condition reduces to
\begin{equation}
\frac{T \vert  \overline{\Delta E } \vert}{V} > 1 \;,
\end{equation}
i.e. the variance of the energy fluctuations must not be too
large. Here we have used the fact that $\overline{\Delta E} < 0$
and the relation $ \overline{e^x} = e^{\overline{x}+
\overline{(x-\overline{x})^2} /2}$ which holds for Gaussian
distributions.

\noindent {\bf Regime II}: When
\begin{equation}
\overline{\left( \frac{w_a^\leftarrow
w_b^\leftarrow}{w_a^\rightarrow w_b^\rightarrow} \right)}  < 1
\leq \overline{\left( \frac{w_a^\leftarrow
w_b^\leftarrow}{w_a^\rightarrow w_b^\rightarrow} \right)^2}
 \;,
\end{equation}
the velocity $v$ is finite but the diffusion constant is infinite.
It can be shown \cite{Bouchaud90} that in this region the long
time behavior is $ \langle x \rangle =vt$ and $\langle x^2 \rangle
- \langle x \rangle^2 \sim t^{2/\mu}$, where $1<\mu<2$. If we
assume a mean value of $\Delta E$ with a Gaussian distribution
about the mean, the condition reduces to
\begin{equation}
1/2 < \frac{T \vert  \overline{\Delta E}  \vert}{V} \leq 1 \;.
\end{equation}
We have again used $ \overline{\Delta E}  < 0$. For this case it
is known \cite{Bouchaud90} that the exponent $\mu$ is given by
$\mu=2 T \vert \overline{\Delta E}  \vert / V$.

\noindent {\bf Regime III}: When
\begin{equation}
\overline{\left( \frac{w_a^\leftarrow
w_b^\leftarrow}{w_a^\rightarrow w_b^\rightarrow} \right)}
> 1 \;,
\end{equation}
the velocity $v$ is zero. More precisely $ \langle x \rangle \sim
t^\mu$ where $\mu < 1$. The diffusion about this drift is
anomalous in the sense that $\langle x^2 \rangle - \langle x
\rangle^2 \sim t^{2 \mu}$. Assuming again a mean value of $\Delta
E$ with a Gaussian distribution about the mean leads to the
condition
\begin{equation}
\frac{T \vert  \overline{\Delta E}  \vert}{V} \leq 1/2 \;,
\end{equation}
where again we have used the fact that $\overline{\Delta E}  < 0$.

\noindent {\bf Sinai diffusion}: When the average bias is exactly
zero,
\begin{equation} \overline{
\log \left( \frac{w_a^\leftarrow w_b^\leftarrow}{w_a^\rightarrow
w_b^\rightarrow} \right) } = 0 \;,
\end{equation}
the system exhibits Sinai diffusion  \cite{Sinai} with $\langle x
\rangle=0$ and $\langle x^2 \rangle \sim (\ln(t/\tau))^4$, where
$\tau$ is the microscopic time needed to move one monomer. Thus,
we are now considering the case $ \overline{\Delta E} =0$.

Note that when
\begin{equation}
\overline{ \log \left( \frac{w_a^\leftarrow
w_b^\leftarrow}{w_a^\rightarrow w_b^\rightarrow} \right) }
> 0 \;,
\end{equation}
namely a reversed bias where $\overline{\Delta E}>0$, similar
regions can be found by interchanging $\rightarrow$ and
$\leftarrow$. For example, when
\begin{equation}
\overline{ \left( \frac{w_a^\rightarrow
w_b^\rightarrow}{w_a^\leftarrow w_b^\leftarrow} \right)^2 } < 1
\;,
\end{equation}
the velocity $v$ and diffusion constant $D$ of the model are
finite. Such results of course require that the molecular motors
remain attached when they reverse direction.

Note also that the three regimes may be identified
\cite{Bouchaud90} according to the parameter $\mu$. In particular,
we identify $\mu
>2$ with regime I, $1<\mu <2$ with regime II, $\mu<1$ with regime III and
$\mu=0$ with Sinai diffusion.

\section{Simple model for the DNA polymerase/exonuclease system}

In this appendix a model of the DNA polymerase/exoneclease system
is studied. It is shown how a more detailed microscopic model than
those studied in the main text also leads to an effective random
forcing energy landscape. However, in contrast to these models,
the location of the transition points into the anomalous dynamics
regime can not be calculated exactly in a straightforward manner.

\begin{figure}
\includegraphics[width=10cm]{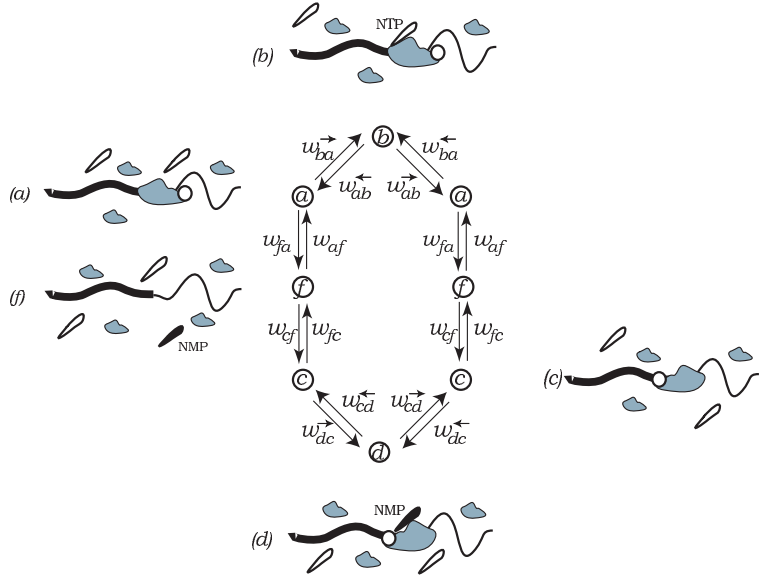} \caption{The possible states
 of the DNA polymerase / exonuclease model. Each pair of either $(a)$,$(f)$ or $(c)$
 states differ by an addition (or removal) of one base from the dsDNA. \label{fig:DNAp}}
\end{figure}

\begin{figure}
\includegraphics[width=10cm]{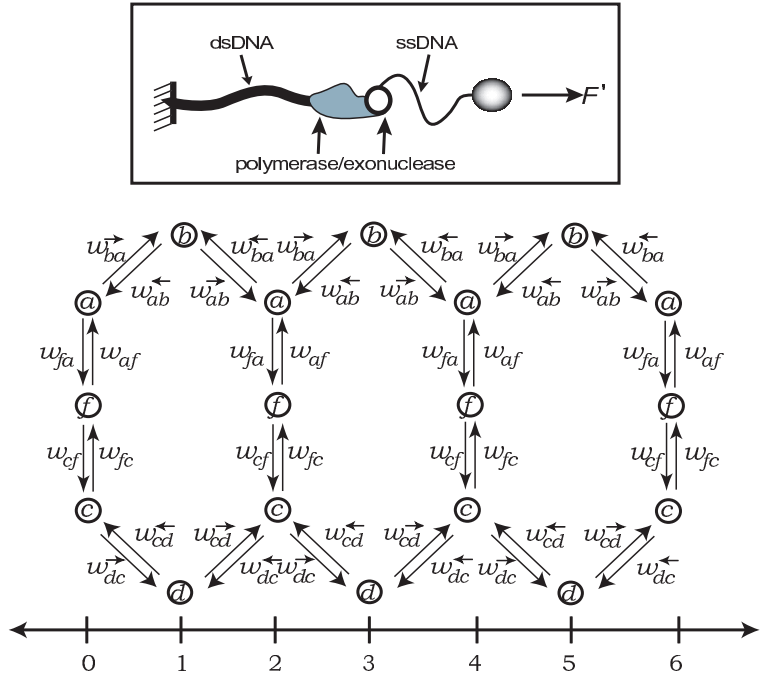} \caption{The full model on the two
lane lattice. The inset on the top depicts a cartoon of the
experimental system. \label{fig:DNApbasic}}
\end{figure}

The model we consider is a simplified version of the model studied
by Goel et al. \cite{Anita03}. The model takes into account the
two active sites of the motor, one acting as a polymerase while
the other acting as an exonuclease. The system can be in one of
five state denoted in Fig. \ref{fig:DNAp} by $(a)$ to $(f)$. The
figure represents only transitions which differ by a motion of the
motor over a distance of one base. The full model along with an
illustration of the experiment is shown in Fig.
\ref{fig:DNApbasic}. In state $(a)$ the motor is attached to the
ssDNA/dsDNA junction with the polymerase active site. In state (b)
the motor uses the energy from the hydrolysis of NTP in order to
be able to extend the dsDNA. States $(c)$ and $(d)$ represent
similar states but now with the motor connected to the junction
using the exonuclease active site. Here the motor does not utilize
energy from the hydrolysis of NTP but instead uses the binding
energy of the NMP. State $(f)$ represents the motor unbound from
the junction. One of the motors in the solution can bind to the
junction in through either the polymerase or exonuclease active
site. Clearly, the model is not a strictly one-dimension model but
corresponds to a random walker moving on two lanes.

The rates of transitions between the states are denoted in the
figure. Explicit expressions similar to Eq. \ref{eq:rates} can
easily be written down. The effect of the external stretching
force $F'$ acting on the ssDNA/dsDNA complex will cause
transitions through the cycle $(a)
\mathop{\rightarrow}\limits^{w_{ba}^\rightarrow} (b)
\mathop{\rightarrow}\limits^{w_{ab}^\rightarrow} (a)$ to be less
favorable while transitions through the cycle $(c)
\mathop{\rightarrow}\limits^{w_{dc}^\leftarrow} (d)
\mathop{\rightarrow}\limits^{w_{cd}^\leftarrow} (c)$ to be more
favorable.
\begin{figure}
\includegraphics[width=8cm]{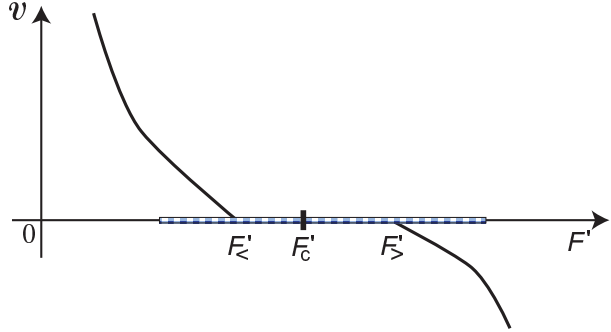} \caption{The expected behavior of the
velocity as a function of the external force $F$ of the DNA
polymerase / exonuclease system. The striped line represents the
region over which the diffusion is expected to be anomalous.
\label{fig:DNApdia}}
\end{figure}

To show that the energy landscape corresponding to the model in
the presence of disorder is indeed a random forcing energy
landscape we first calculate the landscape for the homogeneous
model. Using the results of Derrida \cite{Derrida83} we study one
cycle of the model (see Fig. \ref{fig:DNAp}) and calculate the
ratio of the probabilities $P_a(n)$ and $P_a(n+2)$ of being in the
two $(a)$ states which differ by a translation of one base.
Similarly, the effective energy difference between any two other
sites can be calculated. With the help of Eq. (\ref{eq:derprob})
this ratio can be shown to be given by
\begin{equation}
\frac{P_a(n+2)}{P_a(n)}=\frac{1}{w_{ba}^\leftarrow }\frac{A}{B}
\;, \label{eq:exopoldb}
\end{equation}
with
\begin{eqnarray}
A&=&w_{cf}w_{dc}^\leftarrow w_{cd}^\leftarrow w_{ba}^\leftarrow
w_{ba}^\rightarrow w_{ab}^\rightarrow+w_{af} w_{dc}^\leftarrow
w_{cd}^\leftarrow  w_{ba}^\leftarrow w_{ba}^\rightarrow
w_{ab}^\rightarrow+w_{af} w_{fc} w_{cd}^\leftarrow
 w_{ba}^\leftarrow
w_{ba}^\rightarrow w_{ab}^\rightarrow \nonumber \\
  &+& w_{af} w_{fc} w_{cd}^\rightarrow
w_{ba}^\leftarrow w_{ba}^\rightarrow w_{ab}^\rightarrow  +w_{af}
w_{cd}^\rightarrow w_{dc}^\rightarrow w_{ba}^\leftarrow
w_{ba}^\rightarrow w_{ab}^\rightarrow+w_{cd}^\rightarrow
w_{dc}^\rightarrow w_{cf} w_{ba}^\leftarrow w_{ba}^\rightarrow
w_{ab}^\rightarrow \nonumber \\
&+& w_{cd}^\rightarrow w_{dc}^\rightarrow w_{fa} w_{cf}
 w_{ba}^\rightarrow
w_{ab}^\rightarrow+w_{cd}^\rightarrow w_{dc}^\rightarrow w_{fa}
w_{ab}^\leftarrow w_{cf}  w_{ba}^\leftarrow \;,
\end{eqnarray}
and
\begin{eqnarray}
B&=&w_{ab}^\rightarrow w_{fa} w_{cf} w_{dc}^\leftarrow
w_{cd}^\leftarrow+w_{ab}^\leftarrow w_{fa} w_{cf}
w_{dc}^\leftarrow w_{cd}^\leftarrow+w_{ab}^\leftarrow
w_{ba}^\leftarrow w_{cf} w_{dc}^\leftarrow
w_{cd}^\leftarrow+w_{ab}^\leftarrow w_{ba}^\leftarrow
 w_{af} w_{dc}^\leftarrow w_{cd}^\leftarrow \nonumber \\&+&w_{ab}^\leftarrow w_{ba}^\leftarrow w_{af} w_{fc}
 w_{cd}^\leftarrow+w_{ab}^\leftarrow w_{ba}^\leftarrow w_{af} w_{fc} w_{cd}^\rightarrow
 +w_{ab}^\leftarrow w_{ba}^\leftarrow w_{af} w_{cd}^\rightarrow
w_{dc}^\rightarrow \nonumber \\&+&w_{ab}^\leftarrow
w_{ba}^\leftarrow w_{cd}^\rightarrow w_{dc}^\rightarrow w_{cf} \;.
\end{eqnarray}
The effective energy landscape can be inferred by assuming an
equilibrium distribution so that
\begin{equation}
\frac{P_a(n+2)}{P_a(n)}=\exp((E(n)-E(n+2))/T) \;,
\end{equation}
and the effective energy difference is given by
\begin{equation}
\Delta E = E(n+2)-E(n)=-T \ln\left(\frac{P_a(n+2)}{P_a(n)}\right)
\;.
\end{equation}

It is now clear, using Eq. (\ref{eq:exopoldb}) and arguments
similar to those of Sec. IV, that if the set of rates becomes
site-dependent, a random forcing energy landscape will develop.
The only difference from the simple soluble models studied in the
main text is that the random walker representing the system is
moving on a two-lane lattice. On general grounds \cite{Fisher},
this will not make a difference on the long time and large length
scales behavior of the system. Again, one expects a region when
the velocity is anomalous. The expected behavior of the velocity
as a function of the external force $F'$ is sketched in Fig.
\ref{fig:DNApdia}. Again, we expect that the singularities at
$F'_<$ and $F'_>$ become rounded when $v$ is defined by a finite
experimental time window $t_E$, with a plateau at zero velocity
becoming more and more pronounces as $t_E \to \infty$.

\end{document}